\documentclass[fleqn,usenatbib]{mnras}

\usepackage{newtxtext,newtxmath}

\usepackage[T1]{fontenc}
\usepackage{ae,aecompl}

\usepackage{graphicx}	
\usepackage{amssymb}	

\title[Photon Index Maps of PWNe]{Revealing Hidden Variability in PWNe With Spectral Index Maps}

\author[B. T. Guest \& S. Safi-Harb]{
Benson T. Guest,$^{1,2,3,4}$\thanks{E-mail: umguest@myumanitoba.ca}
Samar Safi-Harb,$^{1}$\thanks{E-mail: samar.safi-harb@umanitoba.ca}
\\
$^{1}$Department of Physics and Astronomy, University of Manitoba, 30a Sifton Rd, Winnipeg R3T 2N2, Canada\\
$^{2}$Department of Astronomy, University of Maryland, College Park, MD 20742, USA\\
$^{3}$X-ray Astrophysics Laboratory,NASA/GSFC, Greenbelt, MD 20771, USA\\
$^{4}$Center for Research and Exploration in Space Science and Technology, NASA/GSFC, Greenbelt, MD 20771, USA\\
}

\date{Accepted XXX. Received YYY; in original form ZZZ}

\pubyear{2020}

\begin{document}
\label{firstpage}
\pagerange{\pageref{firstpage}--\pageref{lastpage}}
\maketitle

\begin{abstract}
Pulsar wind nebulae (PWNe) are the synchrotron bubbles inflated by the rotational energy of a neutron star.
Observing variability within them has previously been limited to cases of significant brightening, or the few instances where transient features are interpreted in terms of intrinsic motion or associated with variability from the pulsar. Jet and torus morphology are also only visible in cases of differing brightness with respect to the surrounding nebula and favourable alignment with our line of sight. Spectral map analysis involves binning observations with an adaptive algorithm to meet a signal limit and colouring the results based on the desired model parameter fits. Minute changes in spectral index become therefore apparent even in cases where brightness images alone do not suggest any underlying changes.
We present a \textit{Chandra} X-ray study of the PWNe in G21.5--0.9, Kes~75, G54.1+0.3, G11.2--0.3, and 3C~58, using archival observations accumulated over the $\sim$20-year lifetime of the mission. With the spectral map analysis technique, we discover evidence for previously unknown variability opening a new window into viewing PWNe.

\end{abstract}



\section{Introduction}
Pulsar wind nebulae (PWNe) are the synchrotron bubbles inflated by the rotational energy of a neutron star (or pulsar). Their study sheds light on their powering engine and the interaction of the relativistic pulsar's wind with its surrounding supernova remnant ejecta or the ISM.
The theory of pulsar winds for many years was assumed to closely follow the work of \cite{KC84a,KC84b} who derived a spherically symmetric MHD model for the evolution, and emission profile for a PWN with synchrotron losses. High resolution X-ray observations have shown that the spectral index ($\alpha, S_{\nu}\sim \nu^{-\alpha}$) in resolved PWNe does not increase as rapidly as predicted by the spherically symmetric model (see e.g., \citet{Guest19} and references therein). Alternative models have been proposed such as diffusion \citep{Tang12}, extending the MHD analysis to 3-D \citep{DelZanna18}, or a combination of both \citep{Porth16}. Spectral maps trace the distribution of particles in both energy and position, and X-ray emitting electrons trace the freshly injected high-energy particles suffering higher synchrotron losses.
Comparison of observed and simulated spectral maps allows an understanding and testing of pulsar wind propagation models.

Observing variability within pulsar wind nebulae has previously been limited to cases of significant brightening, or the few instances where transient features are interpreted in terms of intrinsic motion \citep{Pavlov01,Hester-Variability,2006ApJ...640..929D} or associated with an occasional magnetar-like outburst from the pulsar
\citep{2017ApJ...850L..18B,2016ApJ...824..138Y,Kumar08,2008ApJ...686..508N}. 
In particular, jet and torus morphology have been only visible in cases of differing brightness with respect to the surrounding nebula and favourable alignment with our line of sight. Spectral map analysis involves binning observations with an adaptive algorithm to meet a signal to noise ratio limit and colouring the results based on the desired model parameter fits. Changes in spectral index can become apparent in areas which do not stand out in brightness images alone. In this work, we show that this technique reveals evidence of previously hidden structures and changes in the emitting particle spectrum where traditional RGB and brightness images have provided little insight.

\section{Observations and Methods}

In this work, we focus on archival \textit{Chandra} X-ray observations of relatively young and bright PWNe which have been observed multiple times over the lifetime of the mission.
In Table \ref{tab:ObsList} we summarize our targets (G21.5--0.9, Kes~75, G11.2--0.3, G54.1+0.3 and 3C~58) and the corresponding observations taken with  \textit{Chandra} between its launch in 1999 and the latest observation acquired at the time of this work. We selected these objects given their relative youth and brightness, their dominant non-thermal synchrotron X-ray emission from the PWN, as well as the frequency of archived \textit{Chandra} observations. We note that given their small size, \textit{Chandra} is the most adequate and powerful satellite that allows such a detailed study. 
All observations reported are acquired using the Advanced CCD Imaging Spectrometer (ACIS). Data processing was performed using the Chandra Interactive Analysis of Observations (CIAO) software package \citep{2006SPIE.6270E..1VF}, while spectral analysis made use of the X-ray spectral fitting package XSPEC version 12.9.1 \citep{XSPEC}.

\subsection{The Sample}
\subsubsection{G21.5--0.9}
\begin{figure}
    \centering
    \includegraphics[width=0.95\columnwidth]{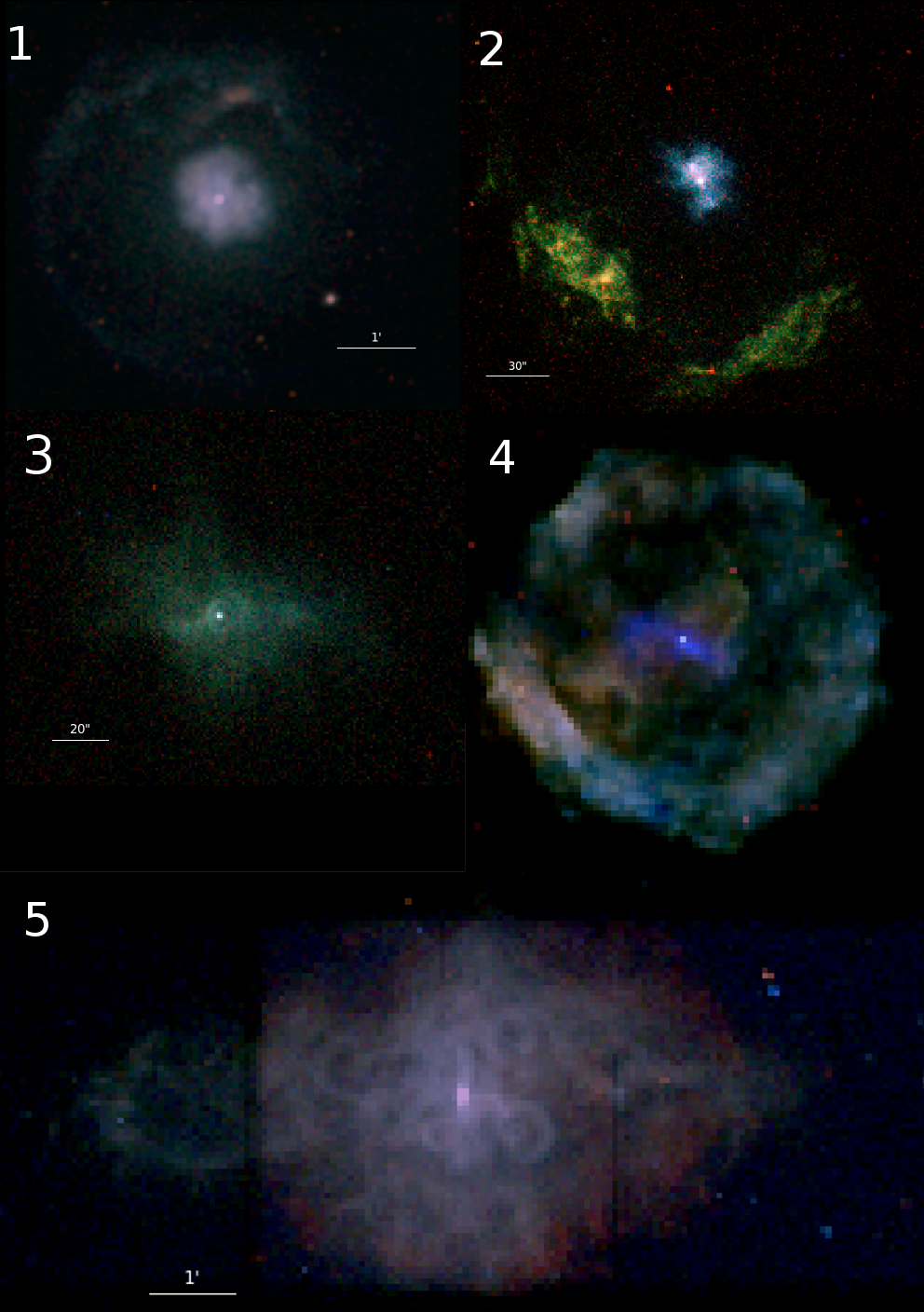}
    \caption[Images of the SNRs observed for spectral map analysis]{Images of the SNRs observed: (1)~G21.5--0.9, (2)~Kes~75, (3)~G54.1+0.3, (4)~G11.2--0.3, and (5)~3C~58. The RGB images are coloured using the energy ranges 0.5--1.2~keV, 1.2--2~keV, and 2--7~keV for red, green, and blue, respectively.}
    \label{fig:SNR-Collection}
\end{figure}

G21.5--0.9 (Figure \ref{fig:SNR-Collection} (1)) is a plerionic composite supernova remnant (SNR) located at a distance of 4.8~kpc \citep{Tian08}. The bright PWN has been used as a calibration target for X-ray missions and has been observed regularly over the 20-year lifetime of the \textit{Chandra X-ray Observatory}. The PWN has a circular morphology extending out to a radius of $\sim 40''$ with steepening spectral index with distance from the centre, and is surrounded by a very faint limb-brightened shell which radiates primarily non-thermal emission \citep{2000ApJ...533L..29S,2001ApJ...561..308S,2005AdSpR..35.1099M}. The only significant source of thermal emission is found from a small knot to the north \citep{Guest19, 2010ApJ...724..572M, 2005A&A...442..539B, 2005AdSpR..35.1099M}. The PWN is powered by the 62~ms pulsar discovered independently in the radio by \cite{Gupta05} and \cite{Camilo06}. \cite{Bietenholz08} measured a PWN expansion age of 870$^{+200} _{-150}$ years using Very Large Array (VLA) observations from 1991 and 2006 making G21.5--0.9 one of the youngest known Galactic PWNe.

\subsubsection{Kes 75}
Kes~75 (Figure \ref{fig:SNR-Collection} (2)) is a plerionic composite SNR at a distance of 5.8~kpc \citep{Verbiest12}. The PWN measures $\sim 26''$ $\times$ 
$20''$ with thermal emission from the supernova remnant shell seen in clumps to the south \citep{Helfand03}. The PWN displays inner structure including a jet directed to the south \citep{Morton07}. The 324 ms pulsar \citep{Gotthelf2000} emitted several magnetar-like outbursts from 2000--2006 \citep{Gavriil08,Kumar08}. This resulted in a softening of the spectrum of the pulsar 
and a brightening of the southern jet from 2000 to 2006 \citep{Kumar08,2008ApJ...686..508N}. \cite{Reynolds18} found a PWN expansion age of $400 \pm 40$ yr making Kes~75 the youngest known Galactic PWN.

\subsubsection{G54.1+0.3}
G54.1+0.3 (Figure \ref{fig:SNR-Collection} (3)) is a plerionic SNR at an estimated distance of 6.3 kpc \citep{Shan18}. The SNR displays a bright point source, surrounding ring, and jet-like elongations to the east and west \citep{Lu02}. While the X-ray emission is all non-thermal, an infrared shell was found by \citep{Temim10} who interpreted the emission as ejecta dust heated by early type stars within the expanding SNR. The 137 ms pulsar powering the nebula was discovered in the radio band by \citep{Camilo02}, who calculated a characteristic age of 2900 years.

\subsubsection{G11.2--0.3}
G11.2--0.3 (Figure \ref{fig:SNR-Collection} (4)) is a plerionic composite SNR \citep{Vasisht96} located at a distance of 5.5--7 kpc \citep{Minter08}). The PWN is powered by a 65~ms pulsar discovered by \citep{Torii97} using \textit{ASCA} X-ray observations. The PWN is elongated with jet-like morphology and broadening at the centre consistent with a torus viewed edge-on \citep{Kaspi01,Roberts03}. The SNR shell displays a circular morphology. \cite{Borkowski16} measured expansion of the shell from 2000--2013. Free expansion places an upper limit of the age as 3600 years however, they argue due to the advanced dynamical age where nearly all the ejecta have been shocked, significant deceleration must have occurred. The unknown density distribution of the surrounding medium leads to age estimates of 1400--2400 years.

\subsubsection{3C~58}
3C~58 (Figure \ref{fig:SNR-Collection} (5)) is a Crab-like SNR possibly associated with the historical supernova event of 1181 C.E. \citep{Stephenson71} and located at a distance of 2 kpc \citep{Kothes13,Kothes16}. The 66 ms pulsar was discovered by \citep{Murray02}. The PWN displays a central jet and torus morphology surrounding the pulsar, the nebula is full of loops and twists, while the outer PWN is embedded in thermal X-ray emission enriched with Ne and Mg suggesting its origin as swept-up ejecta \citep{Slane04,Gotthelf07}.

\begin{table}
    \centering\caption[Observation ID, date, ACIS Chip, and exposure time of the \textit{Chandra} observations used for spectral map analysis]{Observation ID, date, ACIS chip ID and exposure time of the \textit{Chandra} observations used. Data were collected from the Chandra Data Archive (\url{https://cda.harvard.edu/chaser/}).}
    \begin{tabular}{c c c c}
\hline\hline						
Obs ID	&	Date of Obs	&  ACIS Chip  &	Exp Time (ks)		\\\hline
\textbf{G21.5--0.9}	&		&	&		\\
1433	&	1999-11-15	&	S3	&	14.97		\\
1717	&	2000-05-23	&	S3	&	7.54		\\
1725	&	2000-05-24	&	I3	&	7.57		\\
1726	&	2000-05-24	&	I3	&	7.57		\\
1838	&	2000-09-02	&	S3	&	7.85		\\
1839	&	2000-09-02	&	S3	&	7.66		\\
2872	&	2002-09-13	&	I3	&	9.84		\\
2873	&	2002-09-14	&	S3	&	9.83		\\
3699	&	2003-11-09	&	I3	&	9.7		\\
4353	&	2003-05-15	&	S3	&	9.36		\\
5158	&	2005-02-26	&	I3	&	10		\\
5165	&	2004-03-26	&	I3	&	9.55		\\
5166	&	2004-03-14	&	S3	&	10.02		\\
6070	&	2005-02-26	&	I3	&	9.43		\\
6071	&	2005-02-26	&	S3	&	9.64		\\
6740	&	2006-02-21	&	I3	&	9.83		\\
6741	&	2006-02-22	&	S3	&	9.83		\\
8371	&	2007-05-28	&	I3	&	9.92		\\
10644	&	2009-05-29	&	S2	&	9.64		\\
10645	&	2009-05-29	&	S2	&	9.54		\\
10646	&	2009-05-29	&	S3	&	9.54		\\
14263	&	2012-08-08	&	S3	&	9.57		\\
14264	&	2012-08-10	&	I3	&	9.57		\\
16420	&	2014-05-07	&	S3	&	9.57		\\
16421	&	2014-05-09	&	I3	&	9.57		\\
\textbf{Kes 75}	&		&	&		\\
748	&	2000-10-15	& I2  &	37.28		\\
57337	&	2006-06-05  &   S3	&	17.37		\\
6686	&	2006-06-07  &   S3	&	54.1		\\
7338	&	2006-06-09  &   S3	&	39.25		\\
7339	&	2006-06-12  &   S3	&	44.11		\\
10938	&	2009-08-10  &   S3	&	44.61		\\
18030	&	2016-06-08  &   S3	&	84.95		\\
18866	&	2016-06-11  &   S3	&	61.46		\\
\textbf{G11.2--0.3}	&	&	&			\\
780	&	2000-08-06  &   I2	&	19.74		\\
781	&	2000-10-15  &   I2	&	9.97		\\
2322	&	2000-10-15  &   I2	&	4.85		\\
3909	&	2003-05-10  &   I2	&	13.78		\\
3910	&	2003-06-27  &   I2	&	13.78		\\
3911	&	2003-08-01  &   S0	&	14.6		\\
3912	&	2003-09-08  &   I2	&	14.68		\\
14831	&	2013-05-05  &   S3	&	173.02		\\
14830	&	2013-05-25  &   S3	&	58.24		\\
14832	&	2013-05-26  &   S3	&	63.23		\\
15652	&	2013-09-07  &   S3	&	47.93		\\
16323	&	2013-09-08  &   S3	&	45.76		\\
\textbf{G54.1+0.3}	&	&	&			\\
1983	&	2001-06-06  &   I2	&	38.46		\\
9886    &   2008-07-08  &   I2  &   65.33 \\
9108	&	2008-07-10  &   I2	&	34.67		\\
9109	&	2008-07-12  &   I2	&	162.25		\\
9887    &   2008-07-15  &   I2  &   24.84   \\   
\textbf{3C~58}	&	&	&			\\
728	&	2000-09-04  &   S3	&	49.95		\\
3832	&	2003-04-26  &   S3	&	135.83		\\
4382	&	2003-04-23  &   S3	&	167.39		\\
4383	&	2003-04-22  &   S3	&	38.73		\\
\hline
  \end{tabular}  
    \label{tab:ObsList}
\end{table}

\subsection{Methods}
Typically, regions of interest are selected based on the appearance of a brightness or RGB colour image. Here we use the adaptive binning software \textit{CONTBIN} \citep{Contbin} to generate regions following the surface brightness of an input image and meeting a specified signal limit. This limit may be a signal to noise value calculated using a background region, or simply a minimum number of counts. For our analysis where background contribution is relatively small and expected to be uniform across the individual nebulae we use minimum number of counts. This creates puzzle-like pieces that fit together completely covering the area of interest while leaving no gaps and allowing a study of features which may not have been selected for targeted study using handmade regions or that may have been inadvertently averaged over using different region construction methods. Spectra are then extracted from each region and a background taken from outside the PWN, binned to a minimum of 10 counts per bin and fit with the X-ray spectral fitting software \textit{XSPEC} \citep{XSPEC}. We assume that the column density (TBABS model in XSPEC with abundances provided by \cite{Wilms2000}) does not vary over the size of the nebula and freeze this parameter to the best fit value derived from the PWN spectrum. The regions are then coloured according to the values of the desired model parameter to generate maps. When the emission is not purely non-thermal (such as in 3C~58) we can identify the regions where a single component model is insufficient through the construction of a reduced chi-square map and add the thermal component where necessary. Additionally, in cases where there is significant contribution to the surface brightness from the thermal emission associated with the SNR (such as in G11.2--0.3) we restrict our input image and fitting range to the hard X-ray band from 3.5 -- 8 keV where the non-thermal synchrotron emission from the PWN is dominant.

 Pileup occurs when multiple photons are received within the readout time of the detector. These are indistinguishable from a single photon with the sum of the energy. Of our sources, Kes~75, G11.2--0.3 and G54.1+0.3 each has a pulsar which suffers from pile-up. Methods of correcting for pileup require extracting spectra from a region covering the entire point spread function in order to correct for the effect. Our bins surrounding the point source do not meet this requirement, we therefore leave analysis of the central region to traditional methods and other studies, and focus instead on the extended nebular emission which is not piled up.

We investigate changes by creating a merged image to use as the input for region creation. We choose to use a merged image as input to avoid biasing our results towards a single observation. We choose a minimum counts limit such that even the shortest observation will yield a few hundred counts in each region while also attempting to conserve as much inherent structure through keeping the regions as small as possible. Background was taken from regions outside of the remnant while remaining on the same ACIS chip. In each case the background scaled to the individual region sizes is expected to contribute minimally ($\lesssim5$ counts) to the overall total. This combined with our limit of at least $\sim300$ counts per spectrum allows for the use of chi-squared statistics.  Spectra were extracted from the same set of regions for each observation, and those from a common year were fit simultaneously allowing a direct comparison. Significant changes were identified through creating error maps using the same process as above. Regions are coloured according to the 
significance of the difference. This was calculated using the following expression: $(\Gamma_{1}-\Gamma_{2})/\sqrt{\sigma_{1}^{2}+\sigma_{2}^{2}}$,
 where $\Gamma$ is the best-fit photon index for each observation period, and $\sigma$ is the 1-sigma error value associated with each fit. Differences are deemed significant if they exceed the 2-sigma level.
 
 Our analysis introduces several choices which may affect the results. One obvious choice is the signal limit parameter within \textit{Contbin} which alters the number of counts required per region, and therefore defines the resulting region size. The variability we see appears to happen on the scale of a few arcseconds. Increasing the region size much above this has the effect of smoothing out any variability. When we increase the region size sufficiently we no longer see variability and we return to the previous result of PWNe being remarkably stable on large scales. Optimising the region size to produce the best statistics while retaining evidence of variability is complicated by the entanglement of the typical brightness profile of young PWNe and the \textit{Contbin} algorithm. PWNe are brightest in the centre close to the pulsar and decrease in brightness to the edge of the nebula. As a result \textit{Contbin} generates smaller regions close to the pulsar with regions growing larger with distance. In order to examine the effect of region size and statistics we extract spectra from circles of increasing size centred on the region to the north of the pulsar in G21.5-0.9 which showed significant hardening between 2000 and 2005 (Figure \ref{fig:2005-2000-Significance}). For each circle size we compare the difference in photon index and its significance between the 2000 and 2005 observations. We find the significance increases to a maximum for a radius of $3^{\prime\prime}$ and then decreases for larger radii. This roughly matches the size of the \textit{Contbin} region and the number of spectral counts ($\sim 450$) per observation.
A second choice is the freezing of the column density parameter across the nebula.

While we do not expect the column density to vary significantly over the few tens of arcseconds scale of the remnants we studied, we examine the effect of freeing this parameter in our fits of Kes~75. When we look for significant changes between the 2006 and 2016 observations we lose the significant changes in the southern jet, however the region to the north of the pulsar remains as a region of significant softening. 3C58 is an exception as this remnant spans more than 6 arcminutes along its major axis. We note however that the variability we observe occurs in regions near the pulsar spanning $\sim 1^{\prime}$.
A final choice is our threshold for significance. For the number of regions per remnant we have used, it is reasonable to find one or two regions where the observed changes exceed the $2\sigma$ level by chance coincidence. Our analysis relies on the comparison of two or more observation periods. To constrain the errors to a sufficient level to make a meaningful comparison, both observations need to be of sufficient length and quality. While we use the term significant to describe changes detected at a $2\sigma$ level, it should be understood that these changes will need to be confirmed with future pairs of deep observations in order to meet the standard $3\sigma$ detection level.

\subsection{Analysis}
\subsubsection{G21.5--0.9}\label{sec:G21.5-0.9}
\begin{figure}
    \centering
    \includegraphics[width=\columnwidth]{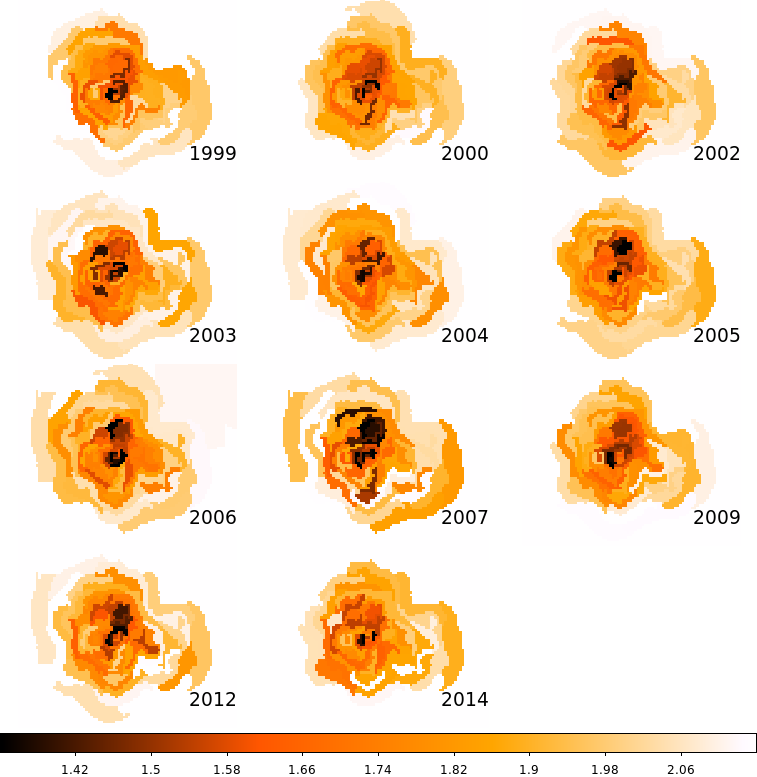}
    \caption[G21.5--0.9 PWN spectral variability maps]{Photon index map of G21.5--0.9. The plume of hard emission to the north of the pulsar does not appear in the first year of observations and becomes more noticeable with time, reaching its hardest spectrum in the February 2005 observations. The plume size appears to change as the surrounding regions also show variability. The maximum size is found in the 2007 observation, however we note the increased uncertainties associated with this observation (Figure \ref{fig:G21-ErrorMap}) compared to the other years.}
    
    \label{fig:G21-Map}
\end{figure}

\begin{figure}
    \centering
    \includegraphics[width=\columnwidth]{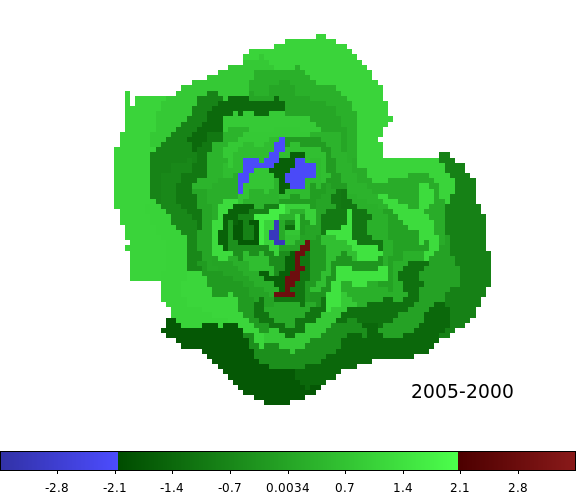}
    \caption[G21.5--0.9 2005 -- 2000 significant difference map]{G21.5--0.9 significance map showing the changes in photon index between observations taken in 2005 and earlier observations from 2000.}
    \label{fig:2005-2000-Significance}
\end{figure}

\begin{figure}
    \centering
    \includegraphics[width=\columnwidth]{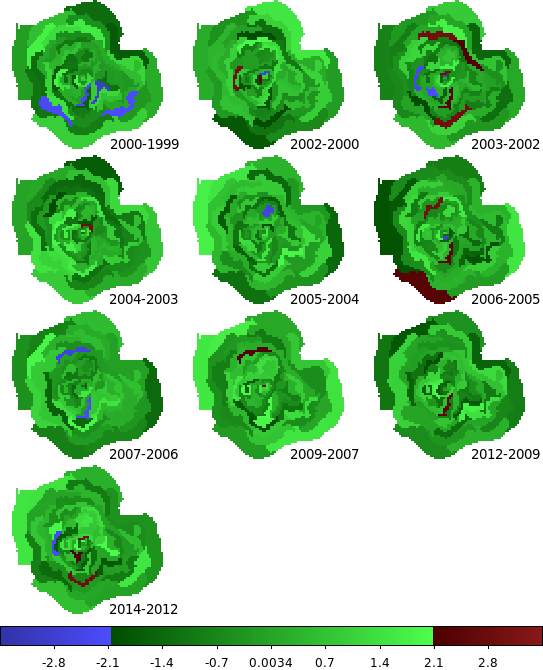}
    \caption[G21.5--0.9 2014 -- 2000 significant difference map]{Significance map of G21.5--0.9 showing the changes in photon index on year-timescales.}
    \label{fig:G21-SignificanceTiles}
\end{figure}

We group observations by year to create individual maps and look for variability in the photon index. Figure \ref{fig:G21-Map} shows the spectral maps by year of observation. Figures \ref{fig:G21-ErrorMap} and \ref{fig:G21-RedChi} show the associated photon index errors and reduced chi-squared values. The region of hard photon index to the north of the pulsar found in the combined map of \cite{Guest19} is not visible in the 1999 observation. It appears hardest in February 2005, and seems to show variability at least on the order of the observation intervals (Table \ref{tab:ObsList}); then hardening again in 2007 and 2012 and softening by the final 2014 observation. To check if these changes are statistically meaningful, we create significance images. 
Figure~\ref{fig:2005-2000-Significance} shows the differences between the 2005 observations where the plume spectrum is the hardest and the 2000 observations. The plume is clearly visible as a region of significant hardening while there is also a region to the south in line with a counter jet that has softened. Figure \ref{fig:G21-SignificanceTiles} shows the significant differences over year timescales.

\begin{figure}
    \centering
    \includegraphics[width=\columnwidth]{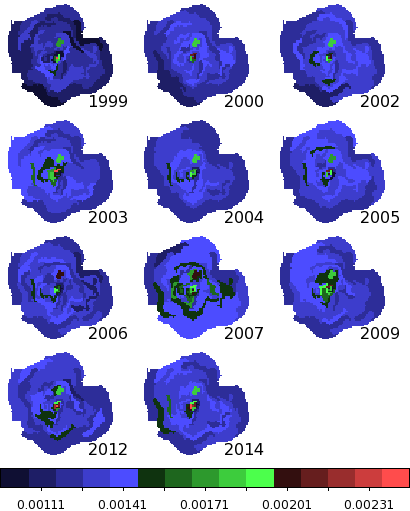}
    \caption[G21.5--0.9 Count rate variability maps]{Exposure corrected flux maps of G21.5--0.9 (in units of photons cm$^{-2}$s$^{-1}$).}
    \label{fig:G21-CountRateMaps}
\end{figure}

A comparison with brightness variability is shown in Figure \ref{fig:G21-CountRateMaps}. Exposure corrected flux images in the broad band (0.5-7 keV) were generated with the \textit{CIAO} scripts \textit{fluximage} for years with a single observation and \textit{merge\_obs} for multiple observation years. The flux images were then binned to match the spectral map regions. The plume to the north of the pulsar shows variability which does not correlate exactly with the changes in photon index. The plume appears brightest in 2006 and 2007 yet is fainter in 2005.

\subsubsection{Kes 75}\label{sec:KES75}
To measure the column density, spectra were extracted from a 35 arcsecond radius centred on the pulsar, with a 5" radius circle excluded to avoid pileup. The spectra were then fit simultaneously with an absorbed power-law with a single column density while allowing the photon index and power-law normalizations to vary between the observation years. The singular column density value found from this fitting was then held constant for each individual region fit in the subsequent spectral maps.

\begin{figure}
    \centering
    \includegraphics[width=\columnwidth]{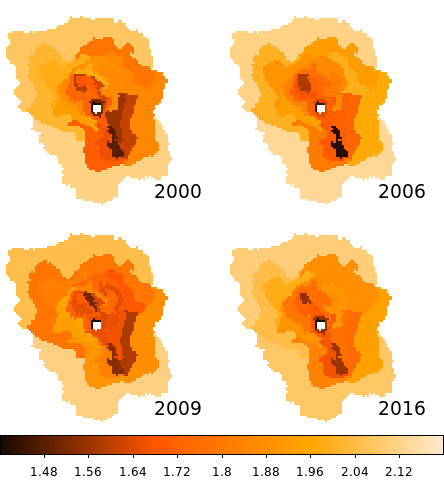}
    \caption[Kes 75 spectral variability map]{Photon index maps vs time for Kes 75. Variability is observed in the jet and surrounding nebula.}
    \label{fig:KES75-SpectralMaps}
\end{figure}

\begin{figure}
    \centering
    \includegraphics[width=\columnwidth]{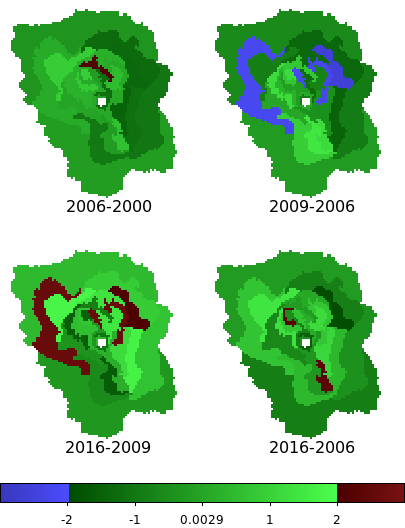}
    \caption[Kes 75 significant difference maps]{Significance maps of Kes~75.  Positive (negative) values correspond to softened (hardened) photon index. The southern jet is visible as a region of softened emission along with a region to the north of the counter jet.}
    \label{fig:Kes75-SignificanceImage}
\end{figure}

Figure \ref{fig:KES75-SpectralMaps} shows the evolution of the Kes~75 spectral maps. Figures \ref{fig:Kes75-ErrorMap} and \ref{fig:Kes75-RedChi} show the associated photon index error and the reduced chi-squared values. Figure \ref{fig:Kes75-FluxImageTiles} shows the exposure corrected flux maps. The southern jet is observed to have the hardest spectrum along with a counter jet in the 2006 observations. In agreement with \citep{2008ApJ...686..508N} the hardest region of the jet does not occur closest to the pulsar, rather there is a gap which shows a softer spectral index which is symmetric around the pulsar. The PWN was observed 4 times for 155 ks over a few days in 2006 and twice for 146 ks in 2016. These observations when fit simultaneously offer tighter constraints than the single observations in 2000 and 2009. We therefore look for significant changes between the 2006 and 2016 observations. Figure \ref{fig:Kes75-SignificanceImage} shows the significance map. There is significant softening in the jet and in a region in the direction of the counter jet, while the rest of the PWN remains largely unchanged. We find hardening between 2006 and 2009 followed by softening between 2009 and 2016 in matching regions. We note the larger errors associated with the 2009 observation (Figure \ref{fig:Kes75-ErrorMap}) compared to the deep observations of 2006 and 2016.

\begin{figure}
    \centering
    \includegraphics[width=\columnwidth]{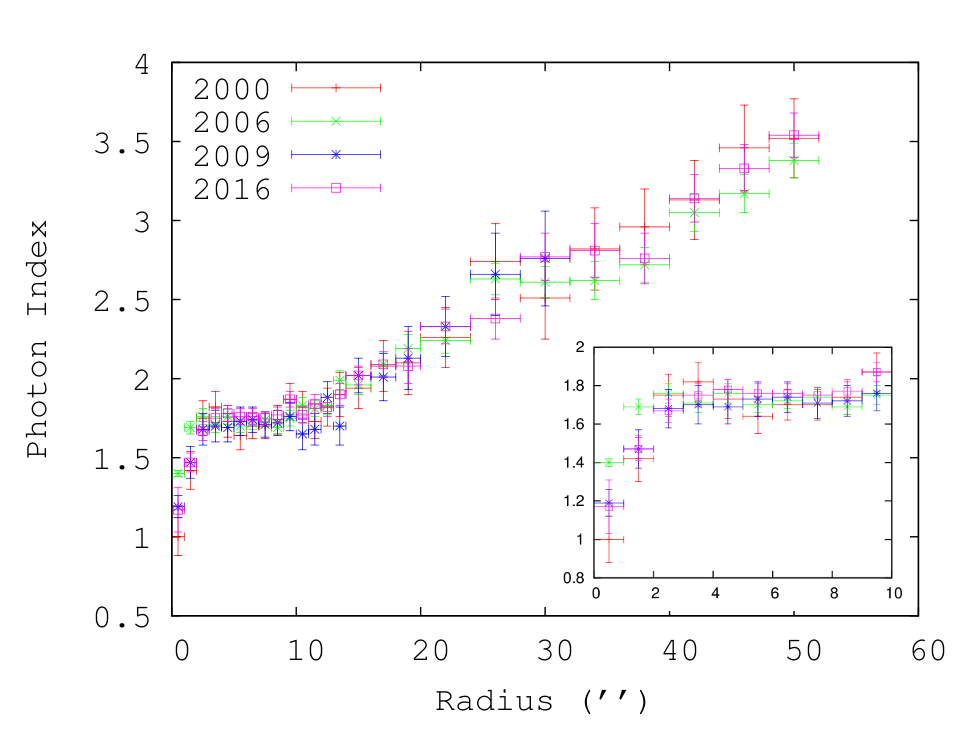}
    \caption[Kes 75 radial profile]{Kes 75 radial spectral profile. The radially averaged spectral index does not change significantly with time over 16 years of observations. The softening of the pulsar emission post outburst is visible however there are no long lasting effects in the nebula. The 2009 profile extends only to $30"$ due to the position on the ACIS detector.}
    \label{fig:KES75-RadialProfile}
\end{figure}

We look for changes in the greater nebula through the radial spectral analysis shown in Figure \ref{fig:KES75-RadialProfile}. Spectra were extracted from rings centred on the pulsar and fit with an absorbed power-law. The central region has been corrected for pile-up. The softening of the pulsar spectrum is observed post outburst in 2006 in agreement with the previous study of \cite{Kumar08}, and returns to the pre-outburst levels by 2009.  Despite the significant changes seen in the spectral maps, the radial profiles do not reveal similar variations between observations in the nebula.

\subsubsection{G54.1+0.3}
\begin{figure}
    \centering
    \includegraphics[width=\columnwidth]{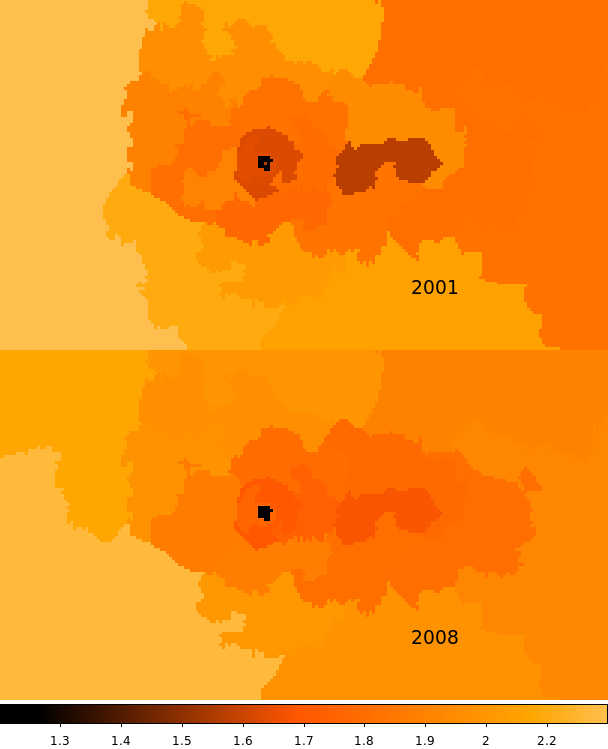}
    \caption{G54.1+0.3 spectral maps. While there appears to be some evidence of softening in the torus and western jet the difference is not considered significant (Figure \ref{fig:G54-SignificanceMap}).}
    \label{fig:G54-SpectralMap}
\end{figure}

\begin{figure}
    \centering
    \includegraphics[width=\columnwidth]{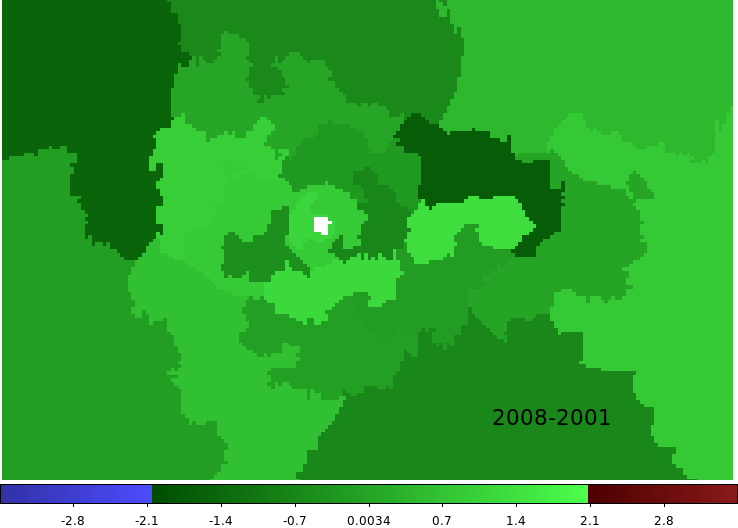}
    \caption[G54.1+0.3 Significance map]{G54.1+0.3 Significance map. We find no significant differences and attribute this to the large errors associated with the brief 2001 observation.}
    \label{fig:G54-SignificanceMap}
\end{figure}

The interesting morphology of G54.1+0.3 with resolved torus and clumpy jets provide an ideal target for spectral map analysis. We see evidence of softening of the torus, most notably on the eastern side and the western jet (Figure \ref{fig:G54-SpectralMap}). Unfortunately, the short observation from 2001 does not allow us to constrain the indices to the degree required to determine if significant changes have occurred (Figures \ref{fig:G54-ErrorMap} and \ref{fig:G54-SignificanceMap}). The reduced chi-squared values are shown in Figure \ref{fig:G54-RedChi}. The exposure corrected flux maps are shown in Figure \ref{fig:G54-ContbinnedFluxMaps}.

\subsubsection{G11.2--0.3}
\begin{figure}
    \centering
    \includegraphics[width=\columnwidth]{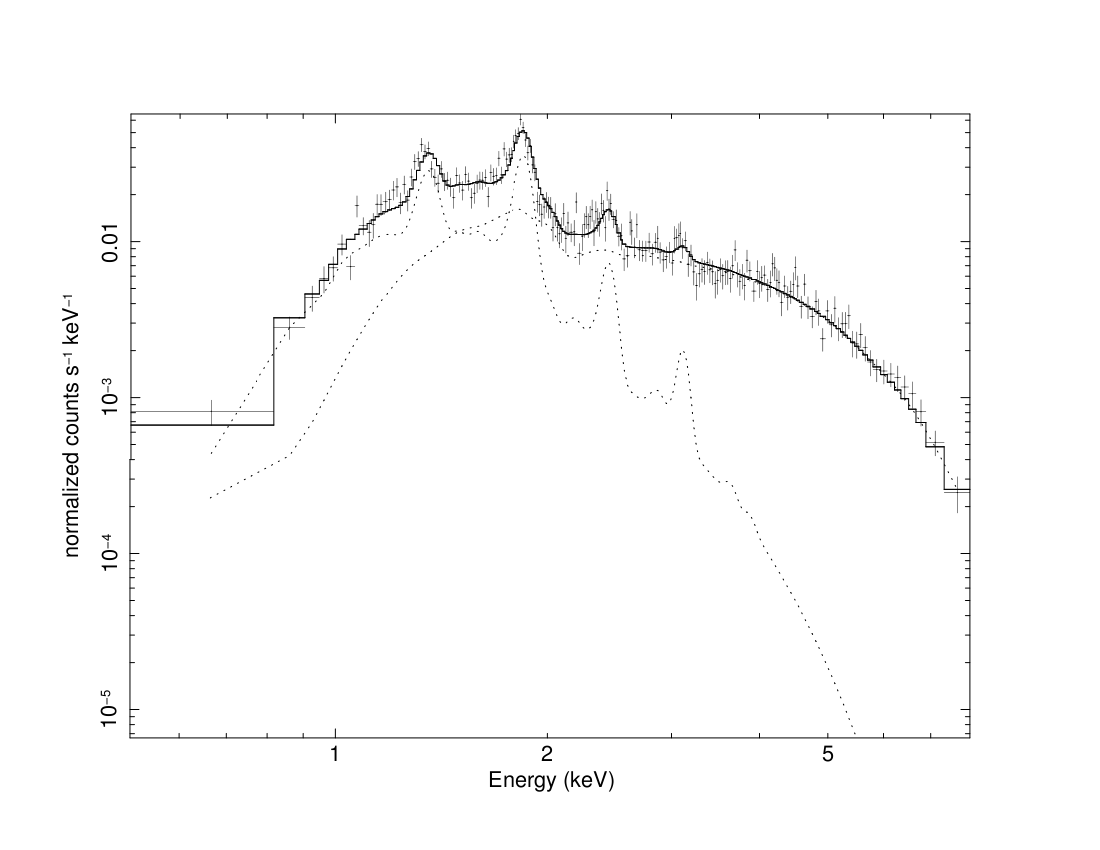}
    \caption[G11.2--0.3 Sample spectrum]{G11.2--0.3 sample spectrum of one of the selected regions displayed with the best fit (solid line) thermal (VPSHOCK) plus non-thermal (Power-law) model. The dotted line shows the components of the model spectrum fitted to the data.}
    \label{fig:G11-SampleSpectrum}
\end{figure}

\begin{figure}
    \centering
    \includegraphics[width=\columnwidth]{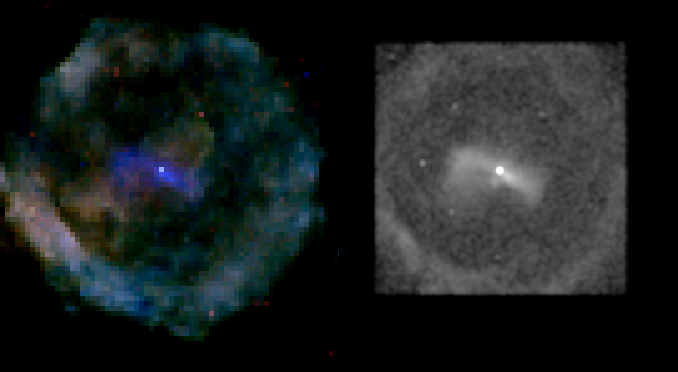}
    \caption[G11.2--0.3 RGB image and hard X-ray image]{Left: RGB \textit{Chandra} Image of G11.2--0.3 with colours defined as 0.5--1.2 keV: red, 1.2--2 keV: green, and 2--7 keV: blue. Right: Filtered 3.5--8 keV image used for spectral map region generation.}
    \label{fig:G11-FilteredImage}
\end{figure}

\begin{figure}
    \centering
    \includegraphics[width=\columnwidth]{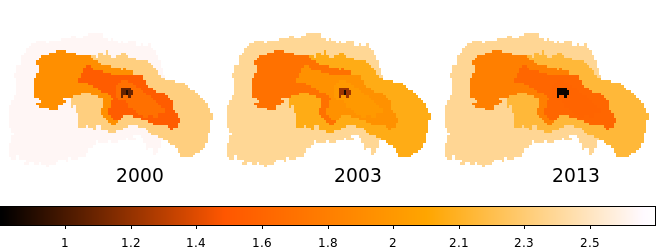}
    \caption[G11.2-0.3 spectral variability maps]{Maps of the photon index values in the PWN of G11.2-0.3}
    \label{fig:G11-SpectralMapZoom}
\end{figure}

The significant thermal emission from the SNR shell fills the remnant and dominates the low energy band with a significant line apparent near 3 keV (Figure \ref{fig:G11-SampleSpectrum}). We therefore filter the input image and spectral fitting range to $3.5-8$ keV to filter out the non-thermal emission from the PWN (Figure \ref{fig:G11-FilteredImage}). The spectral maps appear to show changes with time (Figure \ref{fig:G11-SpectralMapZoom}), however all of the differences are well within the error limits. This is likely due to the relative brevity of the 2000 and 2003 observations leading to large error ranges in the accompanying spectra (Figure \ref{fig:G11-ErrorMap}). Fitting all observations simultaneously to produce a single spectral map merely reproduces the 2013 map which supports this statement. The reduced chi-squared values are shown in Figure \ref{fig:G11-RedChi}. Figure \ref{fig:G11-ContbinnedFluxMaps} shows the exposure corrected flux maps.

\subsubsection{3C~58}
\begin{figure}
    \centering
    \includegraphics[width=\columnwidth]{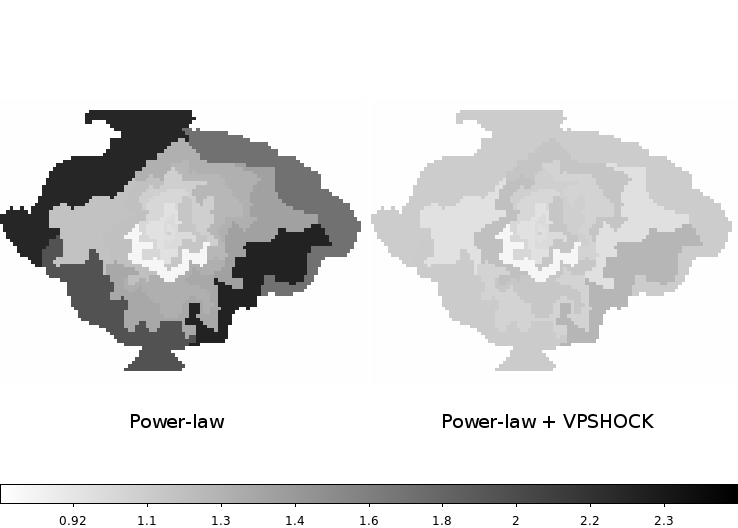}
    \caption[3C~58 reduced chi-squared maps]{3C~58: Maps of the reduced chi-squared value using an absorbed power-law model (left) and adding a vpshock model where required (right).}
    \label{fig:3C58-Chi-Sq}
\end{figure}

\begin{figure}
    \centering
    \includegraphics[width=\columnwidth]{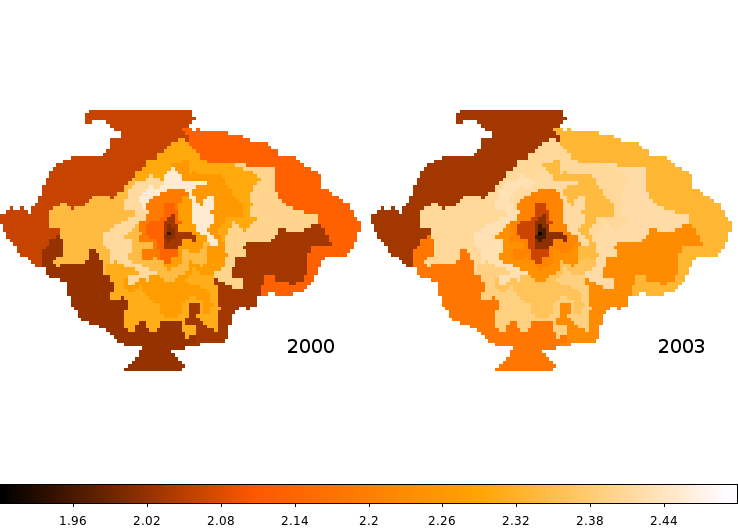}
    \caption[3C~58 spectral variability maps]{Photon index map of 3C~58. The maps are coloured by photon index with darker colours indicating a harder spectrum.}
    \label{fig:3C58-SpectralMap}
\end{figure}

\begin{figure}
    \centering
    \includegraphics[width=\columnwidth]{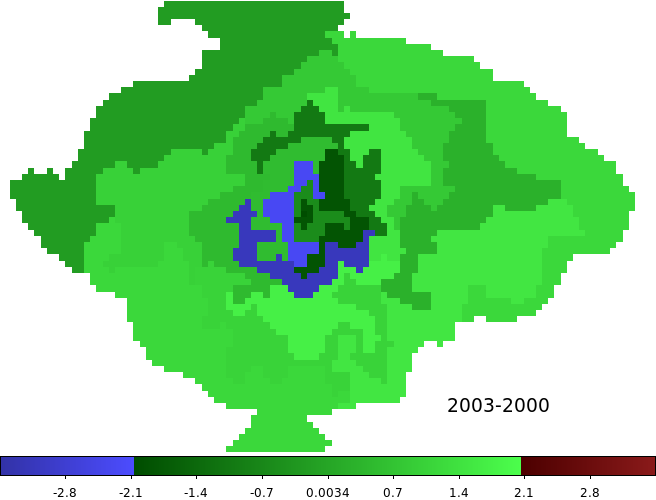}
    \caption[3C58 2003 -- 2000 significant difference map]{
    Significane map of 3C~58 comparing the 2000 to 2003 observations. Positive (negative) values correspond to softened (hardened) photon index between observations. The region surrounding the central pulsar and jet/torus structure has hardened between the observations.}
    \label{fig:3C58-Significance}
\end{figure}

When fit with a single power-law model, the map of reduced chi-squared (Figure \ref{fig:3C58-Chi-Sq}) displays regions where the fit is poor near the edge of the PWN indicating the single power-law model is inadequate. We add a VPSHOCK model to describe the thermal emission which drastically improves the fit. Comparing the spectral maps from 2000 and 2003 (Figure \ref{fig:3C58-SpectralMap}) paired with the significance image (Figure \ref{fig:3C58-Significance}) we find several regions surrounding the pulsar which have significantly hardened over this period. The component error and reduced chi-squared maps with the added two component model regions are shown in Figures \ref{fig:3C58-ErrorMap} and \ref{fig:3C58-RedChi}. Figure \ref{fig:3C58-ContbinnedFluxMaps} shows the exposure corrected flux maps.

\section{Alternative binning methods}
A previous examination of adaptively binned spectral maps was presented in \cite{Kargaltsev17}. Using a weighted Voronoi tesselations (WVT) method they analyze among others the sample of remnants we have studied here. While they simultaneously fit all available observations of each target, their analysis provides an opportunity to contrast the different binning methods. The regions generated by the WVT method do not follow the surface brightness for thin or elongated features such as the southern jet in Kes~75 unless the brightness coupled with the length of observation allows for the regions to be smaller than the observed features. For fainter targets with thin features the WVT method does not utilize the imaging resolution of \textit{Chandra}, as the generated regions effectively smooth these features into their surroundings. For the case of Kes~75 this results in an artificial softening of the southern jet. Our spectral map of Kes~75 shows the jet feature in isolation from its surroundings.

\section{Conclusions}
Spectral maps are a powerful tool which has been under-utilised by the X-ray observing community to study the physics of pulsar wind propagation. We find evidence of variability in G21.5--0.9 in the form of hardening and softening of features resembling an outflow from the pulsar, hardening of the region surrounding the pulsar in 3C~58, hardening of the jets post outburst in Kes 75 followed by softening to return to the pre-outburst state. We find marginal evidence of variability in G11.2--0.3 and G54.1+0.3 which require longer observations to confirm. Examining effects such as the contamination build up on the ACIS detector, we find that this should introduce a systematic hardening which can not explain our observations which show both hardening and softening. We see differences at a $2\sigma$ level using the existing archived observations; more and deeper observations of these and other pulsar wind nebulae are needed to enlarge the sample and improve the statistics to meet the standard $3\sigma$ level. However it is worth noting that we find similarities between previous imaging variability studies and some regions that show variability from our spectral index maps. 

In particular, the plume of variable emission to the north of the pulsar in G21.5--0.9 (Section \ref{sec:G21.5-0.9}) appears in the same location as wisp like features seen in brightness images (\cite{2010ApJ...724..572M,Guest19}). The jets in Kes~75 where we see changes have been observed to vary in brightness (\cite{Reynolds18}). As well as future observations, modelling and comparisons with 3D MHD simulations are required to understand the origins of this newly revealed form of variability. This is beyond the scope of this paper and will be the subject of future follow-up work.

\section*{Acknowledgements}
This research was supported by the Natural Sciences and Engineering Research Council of Canada (NSERC), and made use of the SAO/NASA Astrophysics Data System and the University of Manitoba's High-Energy Catalogue of Supernova Remnants (SNRcat\footnote{http://snrcat.physics.umanitoba.ca/}, \citet{2012AdSpR..49.1313F}). We thank Eric V. Gotthelf and Craig Heinke for comments, and an anonymous referee for a detailed review that helped improve this paper.

\bibliographystyle{mnras}
\bibliography{references}

\begin{thebibliography}{}
\makeatletter
\relax
\def\mn@urlcharsother{\let\do\@makeother \do\$\do\&\do\#\do\^\do\_\do\%\do\~}
\def\mn@doi{\begingroup\mn@urlcharsother \@ifnextchar [ {\mn@doi@}
  {\mn@doi@[]}}
\def\mn@doi@[#1]#2{\def\@tempa{#1}\ifx\@tempa\@empty \href
  {http://dx.doi.org/#2} {doi:#2}\else \href {http://dx.doi.org/#2} {#1}\fi
  \endgroup}
\def\mn@eprint#1#2{\mn@eprint@#1:#2::\@nil}
\def\mn@eprint@arXiv#1{\href {http://arxiv.org/abs/#1} {{\tt arXiv:#1}}}
\def\mn@eprint@dblp#1{\href {http://dblp.uni-trier.de/rec/bibtex/#1.xml}
  {dblp:#1}}
\def\mn@eprint@#1:#2:#3:#4\@nil{\def\@tempa {#1}\def\@tempb {#2}\def\@tempc
  {#3}\ifx \@tempc \@empty \let \@tempc \@tempb \let \@tempb \@tempa \fi \ifx
  \@tempb \@empty \def\@tempb {arXiv}\fi \@ifundefined
  {mn@eprint@\@tempb}{\@tempb:\@tempc}{\expandafter \expandafter \csname
  mn@eprint@\@tempb\endcsname \expandafter{\@tempc}}}

\bibitem[\protect\citeauthoryear{{Arnaud}}{{Arnaud}}{1996}]{XSPEC}
{Arnaud} K.~A.,  1996, in {Jacoby} G.~H.,  {Barnes} J.,  eds,  Astronomical
  Society of the Pacific Conference Series Vol. 101, Astronomical Data Analysis
  Software and Systems V. p.~17

\bibitem[\protect\citeauthoryear{{Bietenholz} \& {Bartel}}{{Bietenholz} \&
  {Bartel}}{2008}]{Bietenholz08}
{Bietenholz} M.~F.,  {Bartel} N.,  2008, \mn@doi [\mnras]
  {10.1111/j.1365-2966.2008.13058.x}, \href
  {https://ui.adsabs.harvard.edu/abs/2008MNRAS.386.1411B} {386, 1411}

\bibitem[\protect\citeauthoryear{{Blumer}, {Safi-Harb}  \&
  {McLaughlin}}{{Blumer} et~al.}{2017}]{2017ApJ...850L..18B}
{Blumer} H.,  {Safi-Harb} S.,   {McLaughlin} M.~A.,  2017, \mn@doi [\apjl]
  {10.3847/2041-8213/aa9844}, \href
  {https://ui.adsabs.harvard.edu/abs/2017ApJ...850L..18B} {850, L18}

\bibitem[\protect\citeauthoryear{{Bocchino}, {van der Swaluw}, {Chevalier}  \&
  {Band iera}}{{Bocchino} et~al.}{2005}]{2005A&A...442..539B}
{Bocchino} F.,  {van der Swaluw} E.,  {Chevalier} R.,   {Band iera} R.,  2005,
  \mn@doi [\aap] {10.1051/0004-6361:20052870}, \href
  {https://ui.adsabs.harvard.edu/abs/2005A&A...442..539B} {442, 539}

\bibitem[\protect\citeauthoryear{{Borkowski}, {Reynolds}  \&
  {Roberts}}{{Borkowski} et~al.}{2016}]{Borkowski16}
{Borkowski} K.~J.,  {Reynolds} S.~P.,   {Roberts} M. S.~E.,  2016, \mn@doi
  [\apj] {10.3847/0004-637X/819/2/160}, \href
  {https://ui.adsabs.harvard.edu/abs/2016ApJ...819..160B} {819, 160}

\bibitem[\protect\citeauthoryear{{Camilo}, {Lorimer}, {Bhat}, {Gotthelf},
  {Halpern}, {Wang}, {Lu}  \& {Mirabal}}{{Camilo} et~al.}{2002}]{Camilo02}
{Camilo} F.,  {Lorimer} D.~R.,  {Bhat} N.~D.~R.,  {Gotthelf} E.~V.,  {Halpern}
  J.~P.,  {Wang} Q.~D.,  {Lu} F.~J.,   {Mirabal} N.,  2002, \mn@doi [\apj]
  {10.1086/342351}, \href
  {https://ui.adsabs.harvard.edu/abs/2002ApJ...574L..71C} {574, L71}

\bibitem[\protect\citeauthoryear{{Camilo}, {Ransom}, {Gaensler}, {Slane},
  {Lorimer}, {Reynolds}, {Manchester}  \& {Murray}}{{Camilo}
  et~al.}{2006}]{Camilo06}
{Camilo} F.,  {Ransom} S.~M.,  {Gaensler} B.~M.,  {Slane} P.~O.,  {Lorimer}
  D.~R.,  {Reynolds} J.,  {Manchester} R.~N.,   {Murray} S.~S.,  2006, \mn@doi
  [\apj] {10.1086/498386}, \href
  {https://ui.adsabs.harvard.edu/abs/2006ApJ...637..456C} {637, 456}

\bibitem[\protect\citeauthoryear{{DeLaney}, {Gaensler}, {Arons}  \&
  {Pivovaroff}}{{DeLaney} et~al.}{2006}]{2006ApJ...640..929D}
{DeLaney} T.,  {Gaensler} B.~M.,  {Arons} J.,   {Pivovaroff} M.~J.,  2006,
  \mn@doi [\apj] {10.1086/500189}, \href
  {https://ui.adsabs.harvard.edu/abs/2006ApJ...640..929D} {640, 929}

\bibitem[\protect\citeauthoryear{{Del Zanna}, {Pili}, {Olmi}, {Bucciantini}  \&
  {Amato}}{{Del Zanna} et~al.}{2018}]{DelZanna18}
{Del Zanna} L.,  {Pili} A.~G.,  {Olmi} B.,  {Bucciantini} N.,   {Amato} E.,
  2018, \mn@doi [Plasma Physics and Controlled Fusion]
  {10.1088/1361-6587/aa9092}, \href
  {http://adsabs.harvard.edu/abs/2018PPCF...60a4027D} {60, 014027}

\bibitem[\protect\citeauthoryear{{Ferrand} \& {Safi-Harb}}{{Ferrand} \&
  {Safi-Harb}}{2012}]{2012AdSpR..49.1313F}
{Ferrand} G.,  {Safi-Harb} S.,  2012, \mn@doi [Advances in Space Research]
  {10.1016/j.asr.2012.02.004}, \href
  {https://ui.adsabs.harvard.edu/abs/2012AdSpR..49.1313F} {49, 1313}

\bibitem[\protect\citeauthoryear{{Fruscione} et~al.,}{{Fruscione}
  et~al.}{2006}]{2006SPIE.6270E..1VF}
{Fruscione} A.,  et~al., 2006, {CIAO: Chandra's data analysis system}.
SPIE, p. 62701V, \mn@doi{10.1117/12.671760}

\bibitem[\protect\citeauthoryear{{Gavriil}, {Gonzalez}, {Gotthelf}, {Kaspi},
  {Livingstone}  \& {Woods}}{{Gavriil} et~al.}{2008}]{Gavriil08}
{Gavriil} F.~P.,  {Gonzalez} M.~E.,  {Gotthelf} E.~V.,  {Kaspi} V.~M.,
  {Livingstone} M.~A.,   {Woods} P.~M.,  2008, \mn@doi [Science]
  {10.1126/science.1153465}, \href
  {https://ui.adsabs.harvard.edu/abs/2008Sci...319.1802G} {319, 1802}

\bibitem[\protect\citeauthoryear{{Gotthelf}, {Vasisht}, {Boylan-Kolchin}  \&
  {Torii}}{{Gotthelf} et~al.}{2000}]{Gotthelf2000}
{Gotthelf} E.~V.,  {Vasisht} G.,  {Boylan-Kolchin} M.,   {Torii} K.,  2000,
  \mn@doi [\apj] {10.1086/312923}, \href
  {https://ui.adsabs.harvard.edu/abs/2000ApJ...542L..37G} {542, L37}

\bibitem[\protect\citeauthoryear{{Gotthelf}, {Helfand}  \&
  {Newburgh}}{{Gotthelf} et~al.}{2007}]{Gotthelf07}
{Gotthelf} E.~V.,  {Helfand} D.~J.,   {Newburgh} L.,  2007, \mn@doi [\apj]
  {10.1086/508767}, \href
  {https://ui.adsabs.harvard.edu/abs/2007ApJ...654..267G} {654, 267}

\bibitem[\protect\citeauthoryear{{Guest}, {Safi-Harb}  \& {Tang}}{{Guest}
  et~al.}{2019}]{Guest19}
{Guest} B.~T.,  {Safi-Harb} S.,   {Tang} X.,  2019, \mn@doi [\mnras]
  {10.1093/mnras/sty2635}, \href
  {http://adsabs.harvard.edu/abs/2019MNRAS.482.1031G} {482, 1031}

\bibitem[\protect\citeauthoryear{{Gupta}, {Mitra}, {Green}  \&
  {Acharyya}}{{Gupta} et~al.}{2005}]{Gupta05}
{Gupta} Y.,  {Mitra} D.,  {Green} D.~A.,   {Acharyya} A.,  2005, Current
  Science, \href {https://ui.adsabs.harvard.edu/abs/2005CSci...89..853G} {89,
  853}

\bibitem[\protect\citeauthoryear{{Helfand}, {Collins}  \& {Gotthelf}}{{Helfand}
  et~al.}{2003}]{Helfand03}
{Helfand} D.~J.,  {Collins} B.~F.,   {Gotthelf} E.~V.,  2003, \mn@doi [\apj]
  {10.1086/344725}, \href
  {https://ui.adsabs.harvard.edu/abs/2003ApJ...582..783H} {582, 783}

\bibitem[\protect\citeauthoryear{{Hester} et~al.,}{{Hester}
  et~al.}{2002}]{Hester-Variability}
{Hester} J.~J.,  et~al., 2002, \mn@doi [\apjl] {10.1086/344132}, \href
  {http://adsabs.harvard.edu/abs/2002ApJ...577L..49H} {577, L49}

\bibitem[\protect\citeauthoryear{{Kargaltsev}, {Klingler}, {Chastain}  \&
  {Pavlov}}{{Kargaltsev} et~al.}{2017}]{Kargaltsev17}
{Kargaltsev} O.,  {Klingler} N.,  {Chastain} S.,   {Pavlov} G.~G.,  2017, in
  Journal of Physics Conference Series. p. 012050 (\mn@eprint {arXiv}
  {1711.02656}), \mn@doi{10.1088/1742-6596/932/1/012050}

\bibitem[\protect\citeauthoryear{{Kaspi}, {Roberts}, {Vasisht}, {Gotthelf},
  {Pivovaroff}  \& {Kawai}}{{Kaspi} et~al.}{2001}]{Kaspi01}
{Kaspi} V.~M.,  {Roberts} M.~E.,  {Vasisht} G.,  {Gotthelf} E.~V.,
  {Pivovaroff} M.,   {Kawai} N.,  2001, \mn@doi [\apj] {10.1086/322515}, \href
  {https://ui.adsabs.harvard.edu/abs/2001ApJ...560..371K} {560, 371}

\bibitem[\protect\citeauthoryear{{Kennel} \& {Coroniti}}{{Kennel} \&
  {Coroniti}}{1984a}]{KC84a}
{Kennel} C.~F.,  {Coroniti} F.~V.,  1984a, \mn@doi [\apj] {10.1086/162356},
  \href {http://adsabs.harvard.edu/abs/1984ApJ...283..694K} {283, 694}

\bibitem[\protect\citeauthoryear{{Kennel} \& {Coroniti}}{{Kennel} \&
  {Coroniti}}{1984b}]{KC84b}
{Kennel} C.~F.,  {Coroniti} F.~V.,  1984b, \mn@doi [\apj] {10.1086/162357},
  \href {http://adsabs.harvard.edu/abs/1984ApJ...283..710K} {283, 710}

\bibitem[\protect\citeauthoryear{{Kothes}}{{Kothes}}{2013}]{Kothes13}
{Kothes} R.,  2013, \mn@doi [\aap] {10.1051/0004-6361/201219839}, \href
  {https://ui.adsabs.harvard.edu/abs/2013A&A...560A..18K} {560, A18}

\bibitem[\protect\citeauthoryear{{Kothes}}{{Kothes}}{2016}]{Kothes16}
{Kothes} R.,  2016, in Supernova Remnants: An Odyssey in Space after Stellar
  Death. p.~46

\bibitem[\protect\citeauthoryear{{Kumar} \& {Safi-Harb}}{{Kumar} \&
  {Safi-Harb}}{2008}]{Kumar08}
{Kumar} H.~S.,  {Safi-Harb} S.,  2008, \mn@doi [\apjl] {10.1086/588284}, \href
  {http://adsabs.harvard.edu/abs/2008ApJ...678L..43K} {678, L43}

\bibitem[\protect\citeauthoryear{{Lu}, {Wang}, {Aschenbach}, {Durouchoux}  \&
  {Song}}{{Lu} et~al.}{2002}]{Lu02}
{Lu} F.~J.,  {Wang} Q.~D.,  {Aschenbach} B.,  {Durouchoux} P.,   {Song} L.~M.,
  2002, \mn@doi [\apjl] {10.1086/340137}, \href
  {http://adsabs.harvard.edu/abs/2002ApJ...568L..49L} {568, L49}

\bibitem[\protect\citeauthoryear{{Matheson} \& {Safi-Harb}}{{Matheson} \&
  {Safi-Harb}}{2005}]{2005AdSpR..35.1099M}
{Matheson} H.,  {Safi-Harb} S.,  2005, \mn@doi [Advances in Space Research]
  {10.1016/j.asr.2005.04.050}, \href
  {https://ui.adsabs.harvard.edu/abs/2005AdSpR..35.1099M} {35, 1099}

\bibitem[\protect\citeauthoryear{{Matheson} \& {Safi-Harb}}{{Matheson} \&
  {Safi-Harb}}{2010}]{2010ApJ...724..572M}
{Matheson} H.,  {Safi-Harb} S.,  2010, \mn@doi [\apj]
  {10.1088/0004-637X/724/1/572}, \href
  {https://ui.adsabs.harvard.edu/abs/2010ApJ...724..572M} {724, 572}

\bibitem[\protect\citeauthoryear{{Minter}, {Camilo}, {Ransom}, {Halpern}  \&
  {Zimmerman}}{{Minter} et~al.}{2008}]{Minter08}
{Minter} A.~H.,  {Camilo} F.,  {Ransom} S.~M.,  {Halpern} J.~P.,   {Zimmerman}
  N.,  2008, \mn@doi [\apj] {10.1086/529005}, \href
  {https://ui.adsabs.harvard.edu/abs/2008ApJ...676.1189M} {676, 1189}

\bibitem[\protect\citeauthoryear{{Morton}, {Slane}, {Borkowski}, {Reynolds},
  {Helfand}, {Gaensler}  \& {Hughes}}{{Morton} et~al.}{2007}]{Morton07}
{Morton} T.~D.,  {Slane} P.,  {Borkowski} K.~J.,  {Reynolds} S.~P.,  {Helfand}
  D.~J.,  {Gaensler} B.~M.,   {Hughes} J.~P.,  2007, \mn@doi [\apj]
  {10.1086/520496}, \href
  {https://ui.adsabs.harvard.edu/abs/2007ApJ...667..219M} {667, 219}

\bibitem[\protect\citeauthoryear{{Murray}, {Slane}, {Seward}, {Ransom}  \&
  {Gaensler}}{{Murray} et~al.}{2002}]{Murray02}
{Murray} S.~S.,  {Slane} P.~O.,  {Seward} F.~D.,  {Ransom} S.~M.,   {Gaensler}
  B.~M.,  2002, \mn@doi [\apj] {10.1086/338766}, \href
  {https://ui.adsabs.harvard.edu/abs/2002ApJ...568..226M} {568, 226}

\bibitem[\protect\citeauthoryear{{Ng}, {Slane}, {Gaensler}  \& {Hughes}}{{Ng}
  et~al.}{2008}]{2008ApJ...686..508N}
{Ng} C.-Y.,  {Slane} P.~O.,  {Gaensler} B.~M.,   {Hughes} J.~P.,  2008, \mn@doi
  [\apj] {10.1086/591146}, \href
  {http://adsabs.harvard.edu/abs/2008ApJ...686..508N} {686, 508}

\bibitem[\protect\citeauthoryear{{Pavlov}, {Kargaltsev}, {Sanwal}  \&
  {Garmire}}{{Pavlov} et~al.}{2001}]{Pavlov01}
{Pavlov} G.~G.,  {Kargaltsev} O.~Y.,  {Sanwal} D.,   {Garmire} G.~P.,  2001,
  \mn@doi [\apjl] {10.1086/321721}, \href
  {http://adsabs.harvard.edu/abs/2001ApJ...554L.189P} {554, L189}

\bibitem[\protect\citeauthoryear{{Porth}, {Vorster}, {Lyutikov}  \&
  {Engelbrecht}}{{Porth} et~al.}{2016}]{Porth16}
{Porth} O.,  {Vorster} M.~J.,  {Lyutikov} M.,   {Engelbrecht} N.~E.,  2016,
  \mn@doi [\mnras] {10.1093/mnras/stw1152}, \href
  {http://adsabs.harvard.edu/abs/2016MNRAS.460.4135P} {460, 4135}

\bibitem[\protect\citeauthoryear{{Reynolds}, {Borkowski}  \&
  {Gwynne}}{{Reynolds} et~al.}{2018}]{Reynolds18}
{Reynolds} S.~P.,  {Borkowski} K.~J.,   {Gwynne} P.~H.,  2018, \mn@doi [\apj]
  {10.3847/1538-4357/aab3d3}, \href
  {http://adsabs.harvard.edu/abs/2018ApJ...856..133R} {856, 133}

\bibitem[\protect\citeauthoryear{{Roberts}, {Tam}, {Kaspi}, {Lyutikov},
  {Vasisht}, {Pivovaroff}, {Gotthelf}  \& {Kawai}}{{Roberts}
  et~al.}{2003}]{Roberts03}
{Roberts} M.~S.~E.,  {Tam} C.~R.,  {Kaspi} V.~M.,  {Lyutikov} M.,  {Vasisht}
  G.,  {Pivovaroff} M.,  {Gotthelf} E.~V.,   {Kawai} N.,  2003, \mn@doi [\apj]
  {10.1086/374266}, \href
  {https://ui.adsabs.harvard.edu/abs/2003ApJ...588..992R} {588, 992}

\bibitem[\protect\citeauthoryear{{Safi-Harb}, {Harrus}, {Petre}, {Pavlov},
  {Koptsevich}  \& {Sanwal}}{{Safi-Harb} et~al.}{2001}]{2001ApJ...561..308S}
{Safi-Harb} S.,  {Harrus} I.~M.,  {Petre} R.,  {Pavlov} G.~G.,  {Koptsevich}
  A.~B.,   {Sanwal} D.,  2001, \mn@doi [\apj] {10.1086/322978}, \href
  {https://ui.adsabs.harvard.edu/abs/2001ApJ...561..308S} {561, 308}

\bibitem[\protect\citeauthoryear{{Sanders}}{{Sanders}}{2006}]{Contbin}
{Sanders} J.~S.,  2006, \mn@doi [\mnras] {10.1111/j.1365-2966.2006.10716.x},
  \href {http://adsabs.harvard.edu/abs/2006MNRAS.371..829S} {371, 829}

\bibitem[\protect\citeauthoryear{{Shan}, {Zhu}, {Tian}, {Zhang}, {Zhang}, {Wu}
  \& {Yang}}{{Shan} et~al.}{2018}]{Shan18}
{Shan} S.~S.,  {Zhu} H.,  {Tian} W.~W.,  {Zhang} M.~F.,  {Zhang} H.~Y.,  {Wu}
  D.,   {Yang} A.~Y.,  2018, \mn@doi [\apjs] {10.3847/1538-4365/aae07a}, \href
  {https://ui.adsabs.harvard.edu/abs/2018ApJS..238...35S} {238, 35}

\bibitem[\protect\citeauthoryear{{Slane}, {Chen}, {Schulz}, {Seward}, {Hughes}
  \& {Gaensler}}{{Slane} et~al.}{2000}]{2000ApJ...533L..29S}
{Slane} P.,  {Chen} Y.,  {Schulz} N.~S.,  {Seward} F.~D.,  {Hughes} J.~P.,
  {Gaensler} B.~M.,  2000, \mn@doi [\apjl] {10.1086/312589}, \href
  {https://ui.adsabs.harvard.edu/abs/2000ApJ...533L..29S} {533, L29}

\bibitem[\protect\citeauthoryear{{Slane}, {Helfand}, {van der Swaluw}  \&
  {Murray}}{{Slane} et~al.}{2004}]{Slane04}
{Slane} P.,  {Helfand} D.~J.,  {van der Swaluw} E.,   {Murray} S.~S.,  2004,
  \mn@doi [\apj] {10.1086/424814}, \href
  {https://ui.adsabs.harvard.edu/abs/2004ApJ...616..403S} {616, 403}

\bibitem[\protect\citeauthoryear{{Stephenson}}{{Stephenson}}{1971}]{Stephenson71}
{Stephenson} F.~R.,  1971, \qjras, \href
  {https://ui.adsabs.harvard.edu/abs/1971QJRAS..12...10S} {12, 10}

\bibitem[\protect\citeauthoryear{{Tang} \& {Chevalier}}{{Tang} \&
  {Chevalier}}{2012}]{Tang12}
{Tang} X.,  {Chevalier} R.~A.,  2012, \mn@doi [\apj]
  {10.1088/0004-637X/752/2/83}, \href
  {http://adsabs.harvard.edu/abs/2012ApJ...752...83T} {752, 83}

\bibitem[\protect\citeauthoryear{{Temim}, {Slane}, {Reynolds}, {Raymond}  \&
  {Borkowski}}{{Temim} et~al.}{2010}]{Temim10}
{Temim} T.,  {Slane} P.,  {Reynolds} S.~P.,  {Raymond} J.~C.,   {Borkowski}
  K.~J.,  2010, \mn@doi [\apj] {10.1088/0004-637X/710/1/309}, \href
  {https://ui.adsabs.harvard.edu/abs/2010ApJ...710..309T} {710, 309}

\bibitem[\protect\citeauthoryear{{Tian} \& {Leahy}}{{Tian} \&
  {Leahy}}{2008}]{Tian08}
{Tian} W.~W.,  {Leahy} D.~A.,  2008, \mn@doi [\mnras]
  {10.1111/j.1745-3933.2008.00557.x}, \href
  {https://ui.adsabs.harvard.edu/abs/2008MNRAS.391L..54T} {391, L54}

\bibitem[\protect\citeauthoryear{{Torii}, {Tsunemi}, {Dotani}  \&
  {Mitsuda}}{{Torii} et~al.}{1997}]{Torii97}
{Torii} K.,  {Tsunemi} H.,  {Dotani} T.,   {Mitsuda} K.,  1997, \mn@doi [\apjl]
  {10.1086/316798}, \href
  {https://ui.adsabs.harvard.edu/abs/1997ApJ...489L.145T} {489, L145}

\bibitem[\protect\citeauthoryear{{Vasisht}, {Aoki}, {Dotani}, {Kulkarni}  \&
  {Nagase}}{{Vasisht} et~al.}{1996}]{Vasisht96}
{Vasisht} G.,  {Aoki} T.,  {Dotani} T.,  {Kulkarni} S.~R.,   {Nagase} F.,
  1996, \mn@doi [\apjl] {10.1086/309854}, \href
  {https://ui.adsabs.harvard.edu/abs/1996ApJ...456L..59V} {456, L59}

\bibitem[\protect\citeauthoryear{{Verbiest}, {Weisberg}, {Chael}, {Lee}  \&
  {Lorimer}}{{Verbiest} et~al.}{2012}]{Verbiest12}
{Verbiest} J.~P.~W.,  {Weisberg} J.~M.,  {Chael} A.~A.,  {Lee} K.~J.,
  {Lorimer} D.~R.,  2012, \mn@doi [\apj] {10.1088/0004-637X/755/1/39}, \href
  {https://ui.adsabs.harvard.edu/abs/2012ApJ...755...39V} {755, 39}

\bibitem[\protect\citeauthoryear{{Wilms}, {Allen}  \& {McCray}}{{Wilms}
  et~al.}{2000}]{Wilms2000}
{Wilms} J.,  {Allen} A.,   {McCray} R.,  2000, \mn@doi [\apj] {10.1086/317016},
  \href {https://ui.adsabs.harvard.edu/abs/2000ApJ...542..914W} {542, 914}

\bibitem[\protect\citeauthoryear{{Younes} et~al.,}{{Younes}
  et~al.}{2016}]{2016ApJ...824..138Y}
{Younes} G.,  et~al., 2016, \mn@doi [\apj] {10.3847/0004-637X/824/2/138}, \href
  {https://ui.adsabs.harvard.edu/abs/2016ApJ...824..138Y} {824, 138}

\makeatother
\end{thebibliography}



\newpage

\appendix

\section{Additional Parameter Maps}

\begin{figure}
    \centering
    \includegraphics[width=\columnwidth]{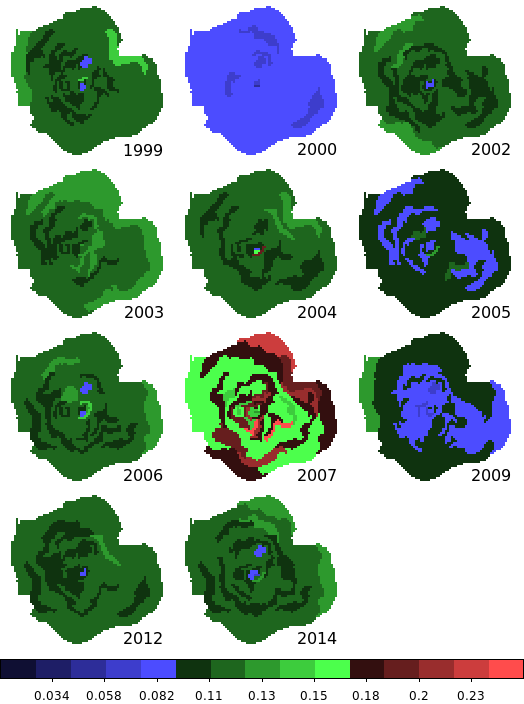}
    \caption{$1-\sigma$ error values of the photon index associated with the fits to G11.2-0.3 shown in Figure \ref{fig:G21-Map}.}
    \label{fig:G21-ErrorMap}
\end{figure}

\begin{figure}
    \centering
    \includegraphics[width=\columnwidth]{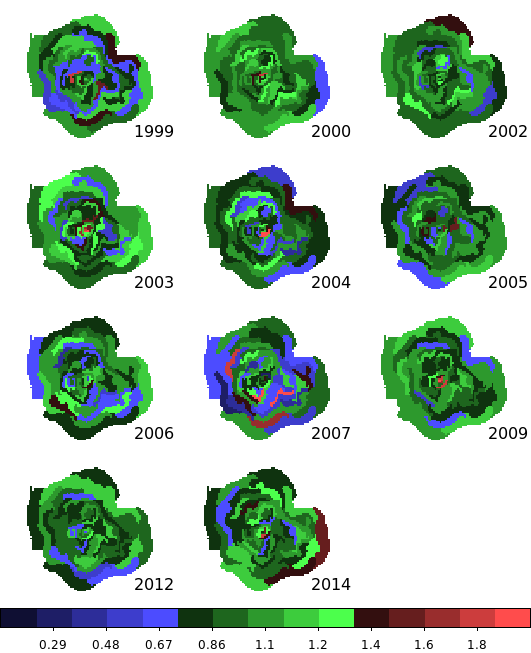}
    \caption{Reduced chi-squared values associated with the fits to G11.2-0.3 shown in Figure \ref{fig:G21-Map}.}
    \label{fig:G21-RedChi}
\end{figure}

\begin{figure}
    \centering
    \includegraphics[width=\columnwidth]{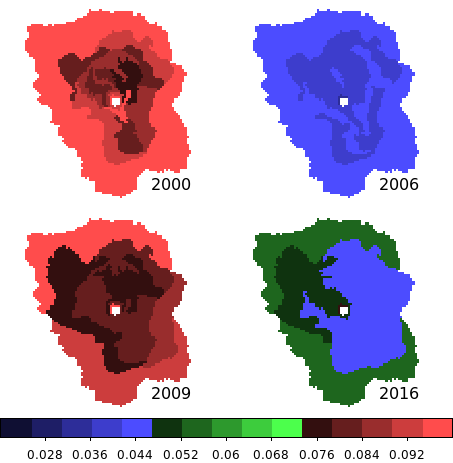}
    \caption{$1-\sigma$ error values in the photon index associated with the fits to Kes~75 shown in Figure \ref{fig:KES75-SpectralMaps}.}
    \label{fig:Kes75-ErrorMap}
\end{figure}

\begin{figure}
    \centering
    \includegraphics[width=\columnwidth]{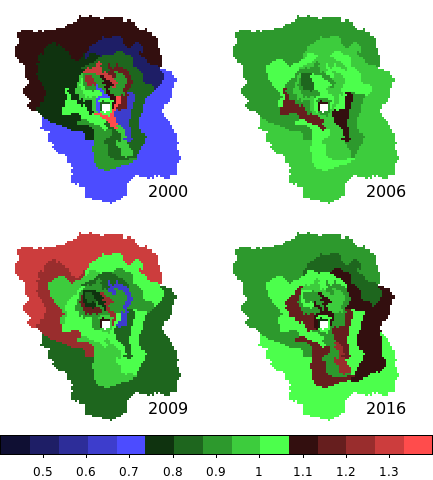}
    \caption{Reduced chi-squared values of the fits to Kes~75 shown in Figure \ref{fig:KES75-SpectralMaps}.}
    \label{fig:Kes75-RedChi}
\end{figure}

\begin{figure}
    \centering
    \includegraphics[width=\columnwidth]{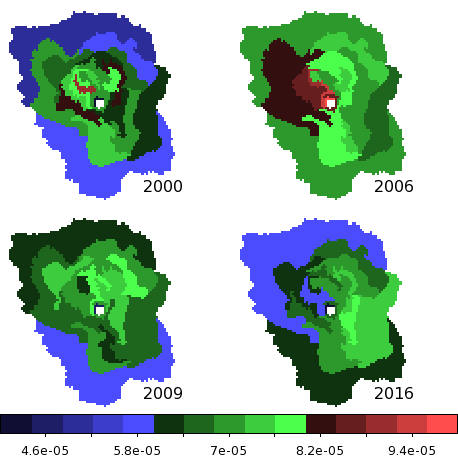}
    \caption{Exposure corrected flux maps of Kes~75 (in units of photons cm$^{-2}$s$^{-1}$).}
    \label{fig:Kes75-FluxImageTiles}
\end{figure}

\begin{figure}
    \centering
    \includegraphics[width=\columnwidth]{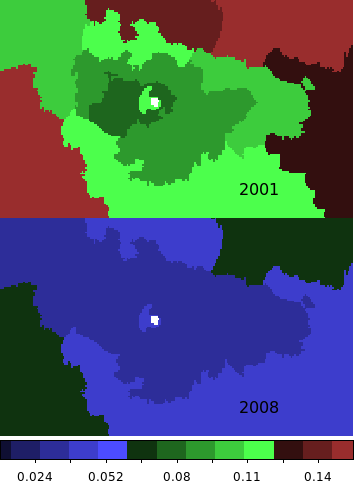}
    \caption{$1-\sigma$ errors of the photon index associated with the fits to G54.1+0.3 shown in Figure \ref{fig:G54-SpectralMap}.}
    \label{fig:G54-ErrorMap}
\end{figure}

\begin{figure}
    \centering
    \includegraphics[width=\columnwidth]{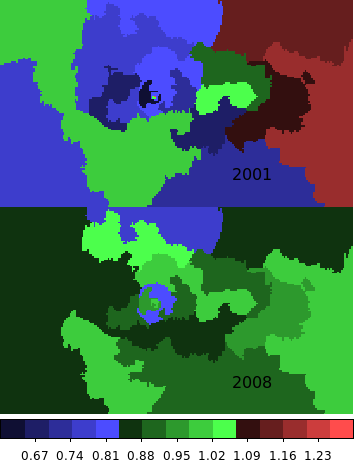}
    \caption{Reduced chi-squared values of the fits to G54.1+0.3 shown in Figure \ref{fig:G54-SpectralMap}.}
    \label{fig:G54-RedChi}
\end{figure}

\begin{figure}
    \centering
    \includegraphics[width=\columnwidth]{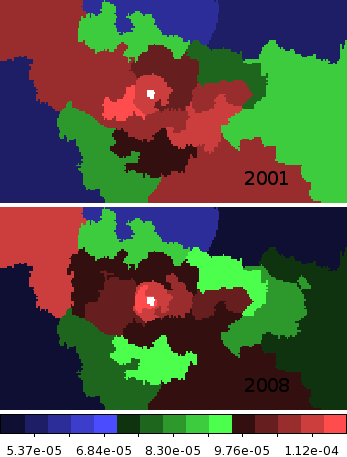}
    \caption{Exposure corrected flux maps of G54.1+0.3 (in units of photons cm$^{-2}$s$^{-1}$).}
    \label{fig:G54-ContbinnedFluxMaps}
\end{figure}

\begin{figure}
    \centering
    \includegraphics[width=\columnwidth]{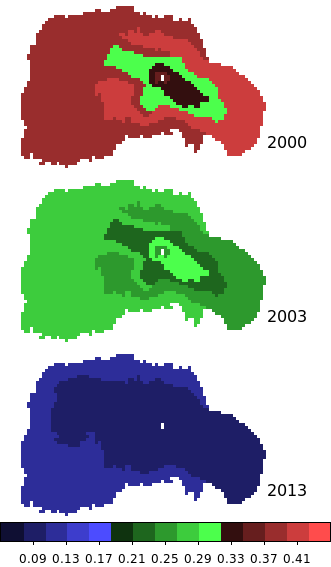}
    \caption{$1-\sigma$ error values of the photon index associated with the fits to G11.2--0.3 shown in Figure \ref{fig:G11-SpectralMapZoom}.}
    \label{fig:G11-ErrorMap}
\end{figure}

\begin{figure}
    \centering
    \includegraphics[width=\columnwidth]{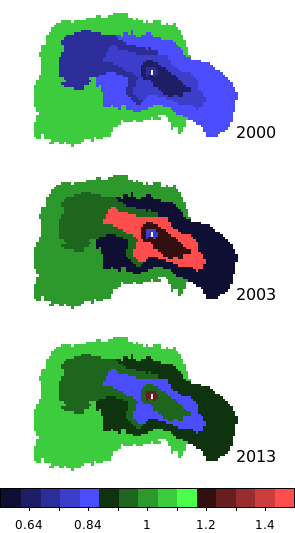}
    \caption{Reduced chi-squared values of the fits to G11.2--0.3 shown in Figure \ref{fig:G11-SpectralMapZoom}.}
    \label{fig:G11-RedChi}
\end{figure}

\begin{figure}
    \centering
    \includegraphics[width=\columnwidth]{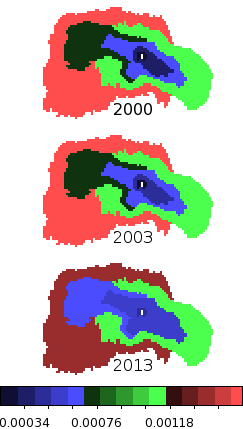}
    \caption{Exposure corrected flux maps of G11.2-0.3 (in units of photons cm$^{-2}$s$^{-1}$).}
    \label{fig:G11-ContbinnedFluxMaps}
\end{figure}

\begin{figure}
    \centering
    \includegraphics[width=\columnwidth]{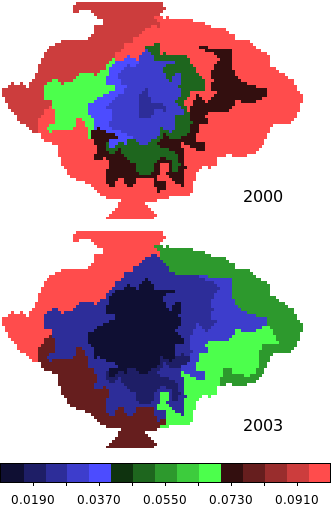}
    \caption{$1-\sigma$ error values of the photon index associated with the fits to 3C58 shown in Figure \ref{fig:3C58-SpectralMap}.}
    \label{fig:3C58-ErrorMap}
\end{figure}

\begin{figure}
    \centering
    \includegraphics[width=\columnwidth]{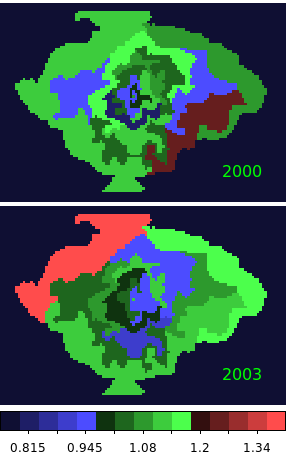}
    \caption{Reduced chi-squared values of the fits to 3C58 shown in Figure \ref{fig:3C58-SpectralMap}.}
    \label{fig:3C58-RedChi}
\end{figure}

\begin{figure}
    \centering
    \includegraphics[width=\columnwidth]{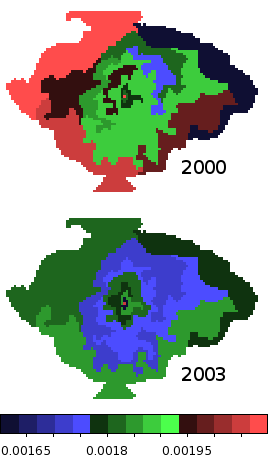}
    \caption{Exposure corrected flux maps of 3C58 (in units of photons cm$^{-2}$s$^{-1}$).}
    \label{fig:3C58-ContbinnedFluxMaps}
\end{figure}


\bsp	
\label{lastpage}
\end{document}